\RequirePackage{amsmath}
\documentclass[runningheads]{llncs}

\usepackage{lipsum}  
\usepackage{ifthen}

\usepackage{graphicx}
\usepackage{subfiles}
\usepackage{float}
\usepackage{amsmath}
\usepackage{subcaption}
\usepackage{amssymb}
\usepackage{multirow}
\usepackage[]{algorithm2e}
\usepackage{tikz}
\usepackage{calc}
\usetikzlibrary{tikzmark,fit}
\usepackage{ifthen}
\usetikzlibrary{shapes,arrows}
\usepackage{booktabs}
\newcommand{\ra}[1]{\renewcommand{\arraystretch}{#1}}
\usepackage{todonotes}
\usepackage{hyperref}
\usepackage{graphicx}
\usepackage{color}
\usepackage[labelformat=simple]{subcaption}

\tikzset{
		pics/convlayer/.style args = {#1/#2/#3/#4/#5}{
		code = {
		    \def\a{#1} % Used to distinguish input resolution for current layer.
			\def\b{0.02}
			\def\c{#2} % Width of the cube to distinguish number of input channels for current layer.
			\def\t{#3} % X offset for current layer.
			\def\d{#4} % Y offset for current layer.

			\draw[line width=0.3mm](\c+\t,0,\d) -- (\c+\t,\a,\d) -- (\t,\a,\d);                                                      % back plane
			\draw[line width=0.3mm](\t,0,\a+\d) -- (\c+\t,0,\a+\d) node[midway,below] {#5} -- (\c+\t,\a,\a+\d) -- (\t,\a,\a+\d) -- (\t,0,\a+\d); % front plane
			\draw[line width=0.3mm](\c+\t,0,\d) -- (\c+\t,0,\a+\d);
			\draw[line width=0.3mm](\c+\t,\a,\d) -- (\c+\t,\a,\a+\d);
			\draw[line width=0.3mm](\t,\a,\d) -- (\t,\a,\a+\d);

			\filldraw[color=white] (\t+\b,\b,\a+\d) -- (\c+\t-\b,\b,\a+\d) -- (\c+\t-\b,\a-\b,\a+\d) -- (\t+\b,\a-\b,\a+\d) -- (\t+\b,\b,\a+\d); % front plane
			\filldraw[color=white] (\t+\b,\a,\a-\b+\d) -- (\c+\t-\b,\a,\a-\b+\d) -- (\c+\t-\b,\a,\b+\d) -- (\t+\b,\a,\b+\d);

			\ifthenelse {\equal{color=white} {}}
			{} % Do not draw colored slice if #4 is blank.
			{\filldraw[color=white] (\c+\t,\b,\a-\b+\d) -- (\c+\t,\b,\b+\d) -- (\c+\t,\a-\b,\b+\d) -- (\c+\t,\a-\b,\a-\b+\d);} % Else, draw a colored slice.

			} 
			}
}

\begin{document}

\newboolean{anonymous}
\setboolean{anonymous}{false}
\title{Registration by tracking for sequential 2D MRI}
%
%\titlerunning{Abbreviated paper title}
% If the paper title is too long for the running head, you can set
% an abbreviated paper title here
%
\ifthenelse
{\boolean{anonymous}}
{\author{Anonymous}
%
% First names are abbreviated in the running head.
% If there are more than two authors, 'et al.' is used.
%
\institute{Anonymous organization, \email{**@******.***}}
}
{
\author{Niklas~Gunnarsson\inst{1,2}\orcidID{0000-0002-9013-949X} \and Jens~Sjölund\inst{2}\orcidID{0000-0002-9099-3522} \and
Thomas~B.~Schön\inst{1}\orcidID{0000-0001-5183-234X}}
\authorrunning{N. Gunnarsson et al.}
% First names are abbreviated in the running head.
% If there are more than two authors, 'et al.' is used.
%
\institute{Department of Information Technology, Uppsala University, Sweden
\email{\{firstname\}.\{surname\}@it.uu.se}\\
 \and
 Elekta Instrument AB, Stockholm, Sweden \\ \email{\{firstname\}.\{surname\}@elekta.com}}
} 

\maketitle              % typeset the header of the contribution

\begin{abstract}
%\todo{real time MRI - Process images in real time -  summary and major benefits of method - result - Fast, modality agnostic, small objects large motions, online learning}

Our anatomy is in constant motion. With modern MR imaging it is possible to record this motion in real-time during an ongoing radiation therapy session. In this paper we present an image registration method that exploits the sequential nature of 2D MR images to estimate the corresponding displacement field. The method employs several discriminative correlation filters that independently track specific points. Together with a sparse-to-dense interpolation scheme we can then estimate of the displacement field. The discriminative correlation filters are trained online, and our method is modality agnostic. For the interpolation scheme we use a neural network with normalized convolutions that is trained using synthetic diffeomorphic displacement fields. The method is evaluated on a segmented cardiac dataset and when compared to two conventional methods we observe an improved performance. This improvement is especially pronounced when it comes to the detection of larger motions of small objects.
\end{abstract}

\section{Introduction}
% General intro to the problem area
Image guided radiation therapy has been a key component to improve the accuracy of radiation therapy \cite{xing2006overview}. Daily imaging of the current anatomical state informs whether to modify the treatment plan or not. With integration of MRI with the treatment machines is today possible to do real-time magnetic resonance imaging (MRI) during a treatment session \cite{paganelli2018mri}. Since MRI is ideally suited for imaging soft tissues, this allows identification of the shape and position of tumors and organs at risk and adjusting of the treatment accordingly. 

In this paper we will describe a new method for medical image registration. Our method is suitable for 2D sequential images and it is inspired by recent progress within the domain of computer vision. We first estimate a sparse displacement field using discriminative correlation filters (DCF)~\cite{danelljan2018learning} to track a fixed number of points in the image sequence. Then, we use a trained sparse-to-dense interpolation scheme in the form of a neural network that uses normalized convolutional layers~\cite{eldesokey2019confidence}. Normalized convolutions~\cite{knutsson1993normalized} use confidence and are well-suited for irregularly sampled data. Advantages of our method include i) it is modality agnostic due to the fact that the model is trained online and customized for the current image type, ii) it allows flexibility in trading off computation versus accuracy due to the freedom in selecting tracking points and iii) it is designed with real-time requirements in mind since each tracker is fast and the trackers are independent of each other. Possible disadvantages might be if a tracker loses its target and the filter is learn online to track a divergent point or if the tracking points are too sparse and movements are missed.

%\subsection{Related methods}
Even though conventional methods have shown good result they are computationally heavy as each registration problem is typically formulated as a large optimization problem that needs to be solved iteratively. Image registration methods can be categorized as physics-based methods, e.g. like Demons~\cite{thirion1998image}, interpolation-based methods or knowledge-based methods \cite{Sotiras2013}. Our method is categorized as an interpolation-based image registration method \cite{Sotiras2013}. The two most reputable interpolation-based techniques are B-splines~\cite{rueckert1999nonrigid} and thin-plate splines (TPS)~\cite{bookstein1989principal}. Revaud et al.\cite{revaud2015epicflow} proposed an interpolation-based method that uses an edge-aware geodesic distance to weight the sample points  when interpolating from a sparse to dense representation.

% Machine learning methods
Recently, the interest for learning-based methods has increased and the results have been impressive~\cite{balakrishnan2019voxelmorph,dalca2019unsupervised}. By training a neural network to predict the solution, the method no longer needs to solve an optimization problem at every time step which drastically reduces the execution time. However, those methods require training data, which directly determines their applicability in the sense that the trained model is specific to the modality of the training data.
\section{Method}
\begin{figure}[ht]
    \centering
    \fbox{
    \begin{tikzpicture}
    \node[text width=3cm] at (-2,0.5) {Images:};
    \node[text width=3cm] (track) at (-2,-1.5) {Tracking:};
    \node[text width=3cm] at (-2,-3) {Sparse-to-dense \\ interpolation:};
    \node[text width=3cm] at (-2,-5.0) {Displacement field};
    
    \node[inner sep=0pt,label={$t=0$}] (img0) at (0,0.5) {\includegraphics[scale=0.15]{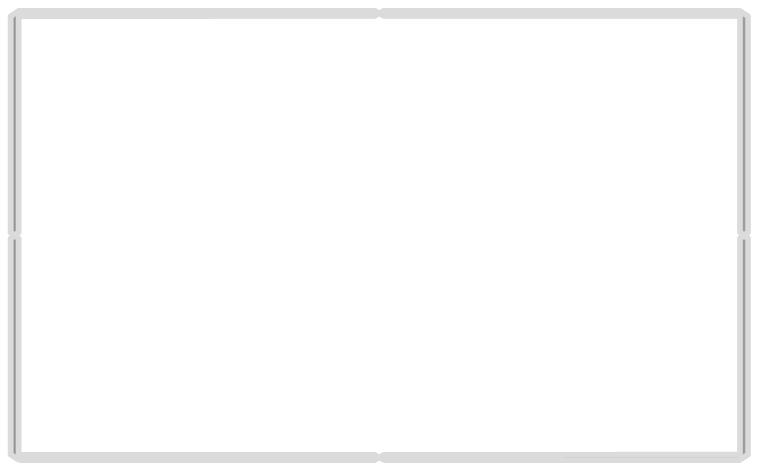}};
    \node[inner sep=0pt,label=below:{Initialization}] (dots0) at (0,-1.5) {\includegraphics[scale=0.15]{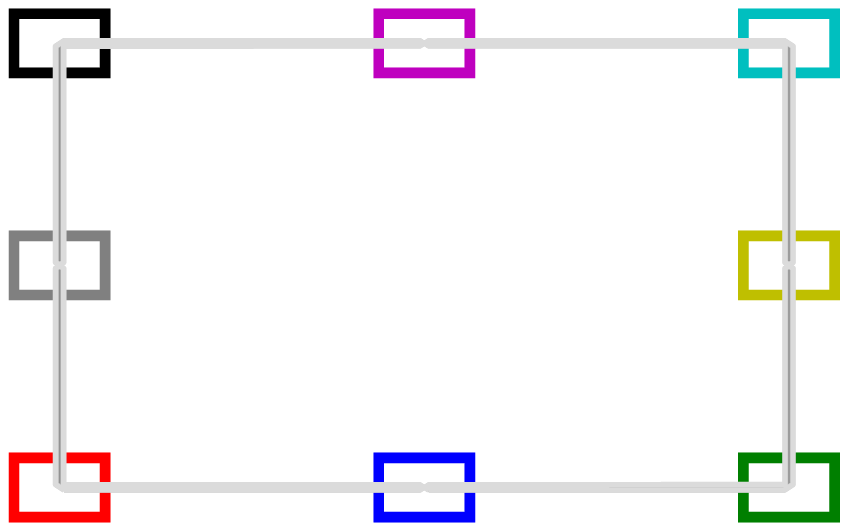}};
    \draw[->,thick] (img0.south) -- (dots0.north);
    \foreach \j [count=\i] in {0,1}
    {
        \node[inner sep=0pt,label={$t=\i$}] (img\i) at (\i * 2,0.5) {\includegraphics[scale=0.15]{Images/workflow/img_\i.eps}};
        \node[inner sep=0pt] (dots\i) at (\i * 2,-1.5){\includegraphics[scale=0.15]{Images/workflow/dots_\i.eps}};
        \node[inner sep=0pt] (sparse\i) at (\i * 2,-3.0) {\includegraphics[scale=0.15]{Images/workflow/sparse_flow_\i.eps}};
        \node[inner sep=0pt] (dense\i) at (\i * 2,-5) {\includegraphics[scale=0.15]{Images/workflow/dense_flow_\i.eps}};
        \draw[->,thick] (img\i.south) -- (dots\i.north);
        \draw[->,thick] (dots\j.east) -- (dots\i.west);
        \draw[->,thick] (dots\i.south) -- (sparse\i.north);
        \draw[->,thick] (sparse\i.south) -- (dense\i.north);
    }
    
    \node[text width=0.5cm](end) at (\i * 2 + 1.5,0.5) {$\hdots$};
    
    \node[draw,inner sep=9pt, rounded corners=0.5cm,fit=(track)(dots0)(sparse\i), label=right:\rotatebox{-90}{\textbf{Suggested method}}](box){};
   
    \end{tikzpicture}
    }
    \caption{From the first image ($t=0$), a fixed number of trackers are initialized at specific points. When a new image ($t=1,2 \dots)$ arrives the tracking positions are updated and a sparse displacement field is estimated. A sparse-to-dense interpolation scheme is used to get the desired dense representation.}
    \label{fig:workflow}
\end{figure}

In the following sections we describe the different steps of the method. We start with the initial location of the trackers, then continue with specific tracker techniques and finally the sparse-to-dense interpolation scheme and how this network is trained. The workflow we suggest is shown in Fig.~\ref{fig:workflow}.

\subsection{Selection of tracking points}
One factor that may have a strong influence on the performance of our method is how the set of tracking points is selected. In computer vision, there has been a range of work on automatically generating candidates such as edges and corners \cite{schmid2000evaluation,shi1994good,stephens1988combined}. On the other hand, it is also possible to use task-based prior knowledge, such as a segmentation mask, when selecting candidates. If the aim is to estimate the deformation of a tumor or organs at risk, a task-based candidate might be preferable considering the fact that the interpolation accuracy tends to deteriorate with the distance~\cite{Sotiras2013}. Therefore, we suggest to sample candidates inversely proportional to the distance from a segmentation mask.
\subsection{Tracking}
%What is tracking in general
%How does it fit into an image registration framework
In an image registration problem, the motion vector from one image to the another is referred to as the displacement field. In conventional image registration methods, the problem is most often formulated as a regression problem where the goal is to find a displacement field that minimizes some metric. A common choice is to use static points in space and estimate the displacement field by the corresponding points in the next image. These points are often referred to as control points. We, on the other hand, define the displacement field from a sparse set of tracking points where the spatial location of these points is constantly updated. In the next section, we describe how to estimate a dense displacement field from the location of the tracking points.

%How does DCFs work (briefly)
We suggest discriminative correlation filters, (DCF) \cite{danelljan2018learning} as trackers which have lately proven extremely useful for visual tracking, achieving state-of-the-art results on several benchmarks \cite{bhat2019learning,bolme2010visual,danelljan2016beyond}. The idea is to train a correlation filter $f$ that distinguishes the tracking target from the background. A continuous score variable $y$ is associated with each image patch $x$ where a low score, $y \approx 0$, represents background while high score, $y \approx 1$, represents the target. The typical choice of the score is a sampled Gaussian function with its mean centered in the target position. The filter is initially trained on a set of image patches, $\{x_i\}_{i=1, \ldots,N}$, wherein the target and the desired output $y_i$ are defined. In practice it is convenient to generate data from the first frame by applying different augmentation techniques, such as rotating, blurring and flipping, etc. In the simple case where the image patch has a single channel the filter is determined by solving a regularized least-squares problem of the form~\cite{henriques2012exploiting}
 \begin{align}
    \label{eq_l2_mosse}
    f^\star = \arg \min_{f} \sum_{i} ||x_i \star f - y_i||^2 + \lambda||f||^2,
\end{align}
but more general versions, including both multi-channel and multi-resolution representations, exist~\cite{danelljan2018learning}.

Once trained, the filter can be used to predict scores by convolving it with subsequent images. For speed, the convolution can be restricted to a smaller search region. For each tracking point, the displacement vector is extracted as the target location corresponding to the highest score. One advantage with DCF is that the filter can easily be updated with new training data (online) without retraining the entire model. 

In this paper, we use the ECO tracker \cite{danelljan2017eco} which uses deep convolutional features together with a dimensionality-reduction step that makes it faster and more robust against overfitting.

\subsection{Sparse-to-dense interpolation}
We suggest a sparse-to-dense interpolation scheme based on normalized convolutions~\cite{knutsson1993normalized}. Traditional convolutions are well-suited for regularly sampled data, but less so if the data is irregularly sampled. Normalized convolution uses a confidence map derived from the data where a value of $0$ indicates no data is given at that point and values close to $1$ indicate a high certainty of the value. A dense signal can then be estimated from a sparse representation based on a non-negative applicability function and some basis functions. Estimating the applicability functions for several normalized convolutional filters using constant basis functions has shown good results~\cite{eldesokey2019confidence,hua2018normalized} when interpolating dense depth scenes from LiDAR data. 

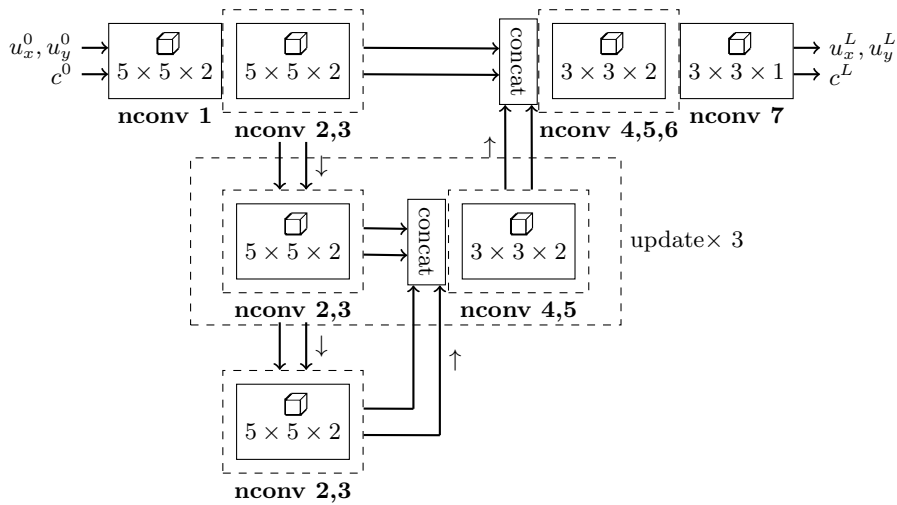
\begin{figure}[ht]
\centering
\begin{tikzpicture}

% level 0
\pic [local bounding box=nconv1] at (0,0) {convlayer = 0.2/0.2/0.0/0.0/$5 \times 5 \times 2$};
\node[draw,inner sep=0pt,minimum height=1cm,fit=(nconv1), label=below:\textbf{nconv 1}](box1){};

\pic [local bounding box=nconv2] at (1.7,0) {convlayer = 0.2/0.2/0.0/0.0/$5 \times 5 \times 2$};
\node[draw,inner sep=0pt,minimum height=1cm,fit=(nconv2)](box2){};
\node[draw, dashed,inner sep=5pt,fit=(box2), label=below:\textbf{nconv 2,3}](update2){};

% level 1
\pic [local bounding box=nconv3] at (1.7,-2.4) {convlayer = 0.2/0.2/0.0/0.0/$5 \times 5 \times 2$};
\node[draw,,inner sep=0pt,minimum height=1cm,fit=(nconv3)](box3){};
\node[draw, dashed,inner sep=5pt,fit=(box3), label=below:\textbf{nconv 2,3}](update3){};

% level 2
\pic [local bounding box=nconv4] at (1.7,-4.8) {convlayer = 0.2/0.2/0.0/0.0/$5 \times 5 \times 2$};
\node[draw,inner sep=0pt,minimum height=1cm,fit=(nconv4)](box4){};
\node[draw, dashed,inner sep=5pt,fit=(box4), label=below:\textbf{nconv 2,3}](update4){};

% level up 1
\node (rect) at (3.5,-2.58) [draw,thin,minimum width=0.5cm,minimum height=1cm](concat1) {\rotatebox{-90}{concat}};
\pic [local bounding box=nconv5] at (4.7,-2.4) {convlayer = 0.2/0.2/0.0/0.0/$3 \times 3 \times 2$};
\node[draw,inner sep=0pt,minimum height=1cm,fit=(nconv5)](box5){};
\node[draw, dashed,inner sep=5pt,fit=(box5), label=below:\textbf{nconv 4,5}](update5){};

% level up 2
\node (rect) at (4.72,-0.18) [draw,thin,minimum width=0.5cm,minimum height=1cm](concat2) {\rotatebox{-90}{concat}};

\pic [local bounding box=nconv6] at (5.9,0) {convlayer = 0.2/0.2/0.0/0.0/$3 \times 3 \times 2$};
\node[draw,inner sep=0pt,minimum height=1cm,fit=(nconv6)](box6){};
\node[draw, dashed,inner sep=5pt,fit=(box6), label=below:\textbf{nconv 4,5,6}](update6){};

\pic [local bounding box=nconv8] at (7.6,0) {convlayer = 0.2/0.2/0.0/0.0/$3 \times 3 \times 1$};
\node[draw,inner sep=0pt,minimum height=1cm,fit=(nconv8), label=below:\textbf{nconv 7}](box8){};

\node[draw, dashed,inner sep=12pt,fit=(update3)(update5), label=right:{update$\times$ 3}](test){};

% arrows
\draw[->,thick] ([xshift=-10, yshift=-5]box1.west) node[left]{$c^0$} -- ([yshift=-5]box1.west) ;
\draw[->,thick] ([xshift=-10, yshift=5]box1.west) node[left]{$u_x^0, u_y^0$} -- ([yshift=5]box1.west);

\draw[->, thick] ([yshift=5]update2.east) -- ([yshift=5]concat2.west);
\draw[->, thick] ([yshift=-5]update2.east) -- ([yshift=-5]concat2.west);

\draw[->,thick] ([xshift=5, yshift=-11]update2.south) -- ([xshift=5]update3.north) node[midway, right](down1){$\downarrow$};
\draw[->,thick] ([xshift=-5, yshift=-11]update2.south) -- ([xshift=-5]update3.north);

\draw[->, thick] ([yshift=5]update3.east) -- ([yshift=5]concat1.west);
\draw[->, thick] ([yshift=-5]update3.east) -- ([yshift=-5]concat1.west);

\draw[->,thick] ([xshift=5, yshift=-11]update3.south) -- ([xshift=5]update4.north) node[midway, right](down2){$\downarrow$};
\draw[->,thick] ([xshift=-5, yshift=-11]update3.south) -- ([xshift=-5]update4.north);

\node (p4) at ([xshift=5, yshift=-60]concat1.south){};
\node (p5) at ([xshift=-5, yshift=-50]concat1.south){};

\draw[->,thick] (p4.north) -- ([xshift=5]concat1.south) node[midway, right](down2){$\uparrow$};
\draw[->,thick] (p5.north) -- ([xshift=-5]concat1.south);

\draw[-,thick] ([yshift=-5]update4.east) -- (p4.north);
\draw[-,thick] ([yshift=5]update4.east) -- (p5.north);

\draw[->,thick] ([xshift=-5]update5.north) -- ([xshift=-5]concat2.south) node[midway, left](down2){$\uparrow$};
\draw[->,thick] ([xshift=5]update5.north)  -- ([xshift=5]concat2.south);

\draw[->,thick] ([yshift=-5]box8.east) -- ([xshift=10, yshift=-5]box8.east) node[right]{$c^L$};
\draw[->,thick] ([yshift=5]box8.east) -- ([xshift=10, yshift=5]box8.east) node[right]{$u_x^L, u_y^L$};

\end{tikzpicture}
    \caption{An illustration of our sparse-to-dense interpolation network. As input we use the channels of the sparse displacement field ($u_x^0$ and $u_y^0$) and a confidence map ($c^0$) and as output we use the dense displacement field for each channel ($u_x^L$, $u_y^L$) and the propagated confidence map ($c^L$). The network consists of seven normalized convolutional layers where layer 2,3,4 and 5 are reused at different feature levels. Each layer is illustrated with a solid box and the numbers indicate $\text{kernel height} \times \text{kernel width} \times \text{number of filters}$. Several subsequent filters are indicated with a dashed line and the filter names around the box. $\uparrow$ and $\downarrow$ are upsampling and downsampling~\cite{eldesokey2019confidence} operators.}
    \label{fig:Nconv}
\end{figure}

We design our interpolation scheme as a neural network inspired by U-net~\cite{ronneberger2015u} with normalized convolutional layers. A detailed illustration of the network is shown in Fig.~\ref{fig:Nconv}. We propagate the sparse displacement field and the confidence map through the network by using non-negative constraints on the weights and as a loss function we use a combination of Huber loss, to minimize the estimation error, and a term that maximizes the output confidence~\cite{eldesokey2019confidence}. 

We generate synthetic displacement fields to train our model. We assume that the displacement field is diffeomorphic and use a geodesic shooting method~\cite{miller2006geodesic} to  generate synthetic displacement fields independently of any imaging data. We artificially create a sparse representation by uniformly sampling coordinates in the generated displacement field. We use a binary confidence map based on the sampled coordinates. To avoid overfitting, we generate unique displacement fields for every iteration. The model is trained offline and only executed when interpolating the sparse representation from the trackers to the dense displacement field. 
\section{Experiments}
For evaluation we used a publicly available cardiac MRI dataset\footnote{\url{http://www.cse.yorku.ca/~mridataset/}}. The dataset contains measurements from 32 subjects (one was removed due to poor image quality). The data of each subject contains an image sequence of 20 images where the ventricle starts in an expanded position, after a few time steps the ventricle has contract an then expands again to its starting position. Each time step has $8$ to $15$ slices of $256 \times 256$ pixels and the slices include manual segmentation of the epicardium and endocardium. We used the manual segmentation masks to segment the myocardium (region in between the epicardium and endocardium) and the ventricle (region inside the endocardium). Since our method is 2D based we select a specific slice in the dataset. We compared our method with several others by applying the estimated displacement fields on the segmentation and calculating the Dice score~\cite{dice1945measures}, 
\begin{align}
    \text{Dice} = \dfrac{2 |S_{t} \cap \hat{S}_{t_0 \to t} |}{|S_t| + |\hat{S}_{t_0 \to t}|}
\end{align}
where $S_t$ is the manual segmentation at time $t$ and $\hat{S}_{t_0 \to t}$ is the warped segmentation using the image at time step $t_0$ and the estimated displacement field between the image at time step $t_0$ and the image at time step $t$. If this paper is accepted the implementation will be publicly available here\footnote{\ifthenelse
{\boolean{anonymous}}
{***}{\url{https://github.com/}}}.
\subsection{Setup}
We initialized 150 different trackers by sampling points inversely proportional to the distance of the union of the two segmentation masks. For each tracker we used an image patch of size $10 \times 10$ pixels and a search area of $4.5$ times the patch size. The trackers were trained on the first image and the filter coefficients are updated online after every new image. For the sparse-to-dense interpolation scheme we used a binary confidence map with $1$ for tracking coordinates and $0$ elsewhere.
\subsection{Result}
For a specific image plane ($z = 4$) the Dice scores between the warped and ground truth segmentation masks were calculated. For comparison we used the following methods i) the same tracker (ECO) but replacing the spare-to-dense interpolation scheme with thin-plate splines (TPS)~\cite{bookstein1989principal}, ii) B-spline image registration~\cite{klein2007evaluation} with mean square error using a grid size of $20 \times 20$ pixels and iii) the Demons algorithm~\cite{thirion1998image}. Table~\ref{tab_result} shows the mean and standard deviation of the result for all image frames and subjects.
\begin{table*}
\centering
\caption{The result for all images and subjects. The Dice mean ($\mu$) and standard deviation ($\sigma$) is shown for each method and segmentation regions (myocardium and ventricle).}
\label{tab_result}
\ra{1.3}
\begin{tabular}{@{}lrrrrcrrrcrrr@{}}\toprule
& \phantom{abc} & \multicolumn{2}{c}{Myocardium (Dice)} & \phantom{abc} && \multicolumn{2}{c}{Ventricle (Dice)} &
\phantom{abc} \\
\cmidrule{3-4} \cmidrule{6-8}
& & $\mu$ & $\sigma$ &&& $\mu$ & $\sigma$\\ \midrule
\textbf{ECO + NCONV} & & $0.810$ & $0.066$ &&& $0.870$ & $0.086$\\

ECO + TPS\cite{bookstein1989principal} & & $0.791$ & $0.062$ &&& $0.866$ & $0.081$\\

B-Spline\cite{klein2007evaluation} & & $0.730$ & $0.148$ &&& $0.837$ & $0.125$\\

Demons\cite{thirion1998image} & & $0.760$ & $0.111$ &&& $0.853$ & $0.111$\\
\bottomrule
\end{tabular}
\end{table*}

\begin{figure}
    \centering
    \includegraphics[width=.75\textwidth]{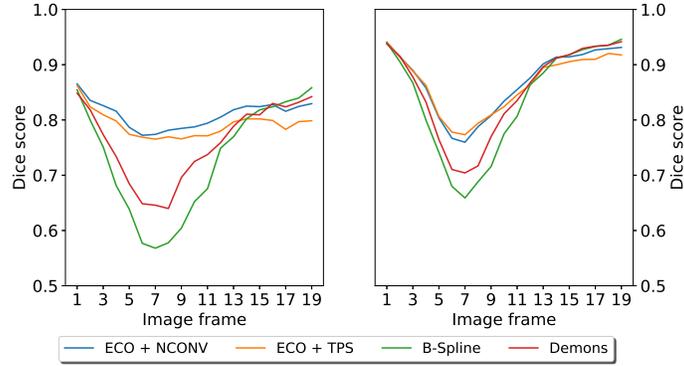}
    \caption{The figure shows the mean Dice score at each time step, averaged across all subjects for both the myocardium (left) and ventricle (right) segmented areas.}
    \label{fig:average}
\end{figure}

\begin{figure}
\centering
\begin{subfigure}{.20\textwidth}
  \centering
  \includegraphics[width=1.0\linewidth]{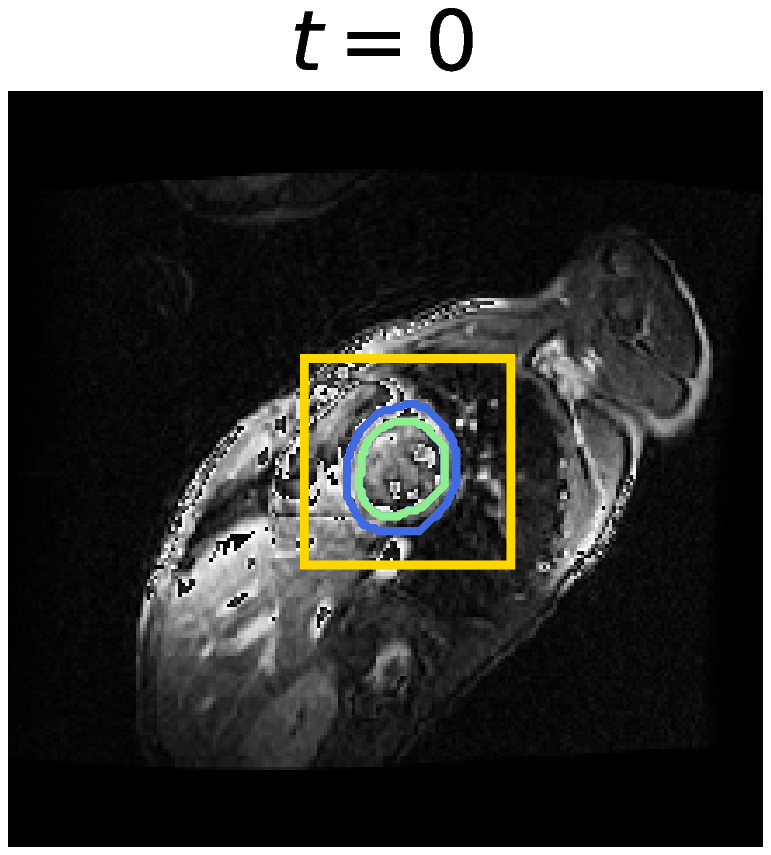}  
  \caption{}
  \label{fig:sub-mov}
\end{subfigure}
\begin{subfigure}{.20\textwidth}
  \centering
  \includegraphics[width=1.0\linewidth]{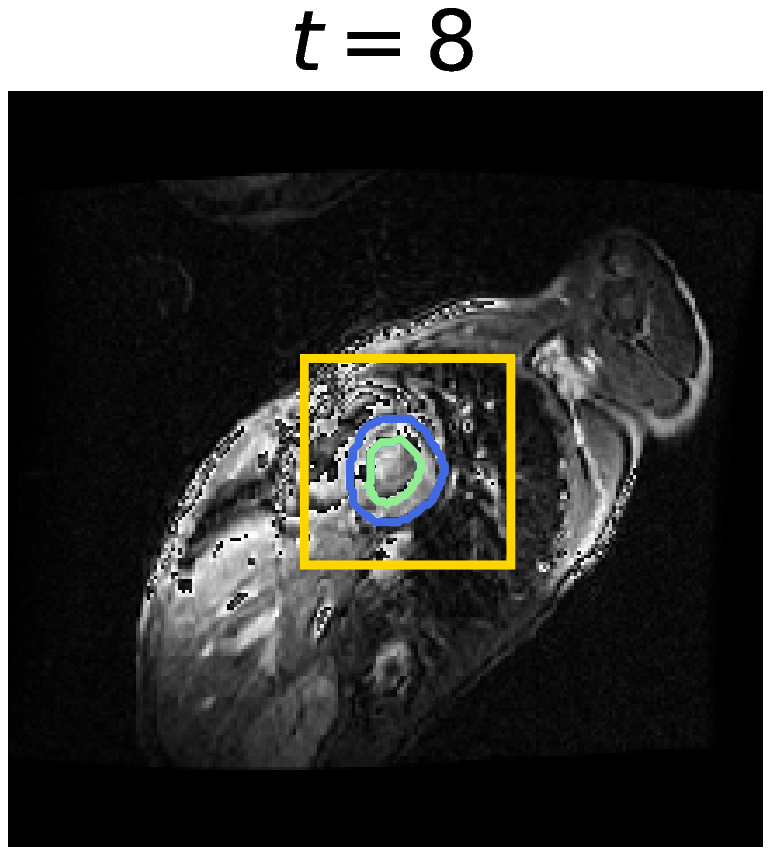}  
  \caption{}
  \label{fig:sub-fixed}
\end{subfigure}
\\
\begin{subfigure}{.19\textwidth}
  \centering
  \includegraphics[width=1.0\linewidth]{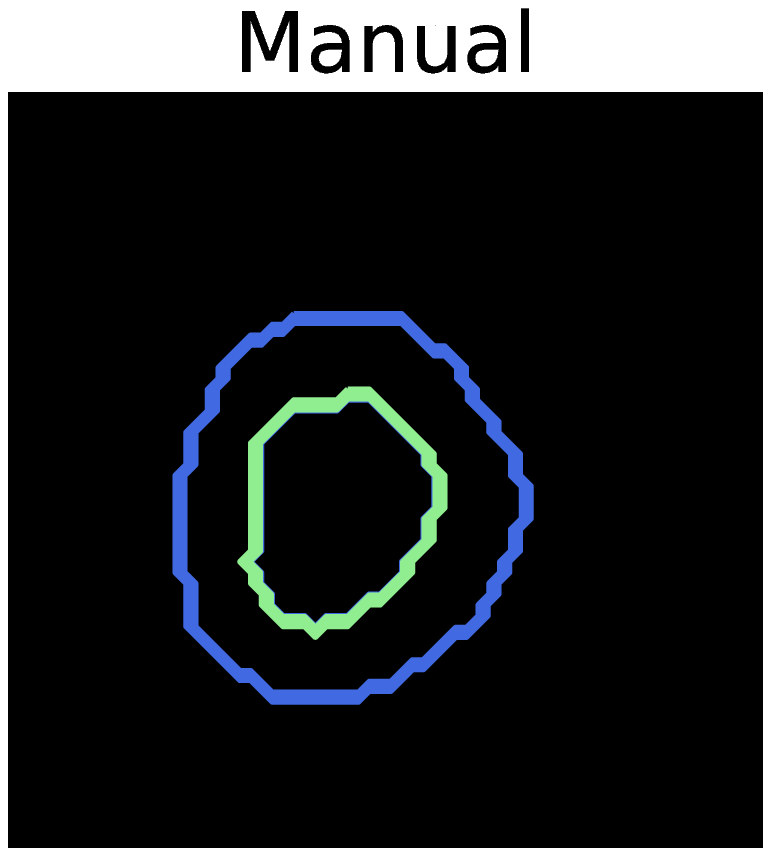}  
  \caption{}
  \label{fig:sub-manual}
\end{subfigure}
\begin{subfigure}{.19\textwidth}
  \centering
  \includegraphics[width=1.0\linewidth]{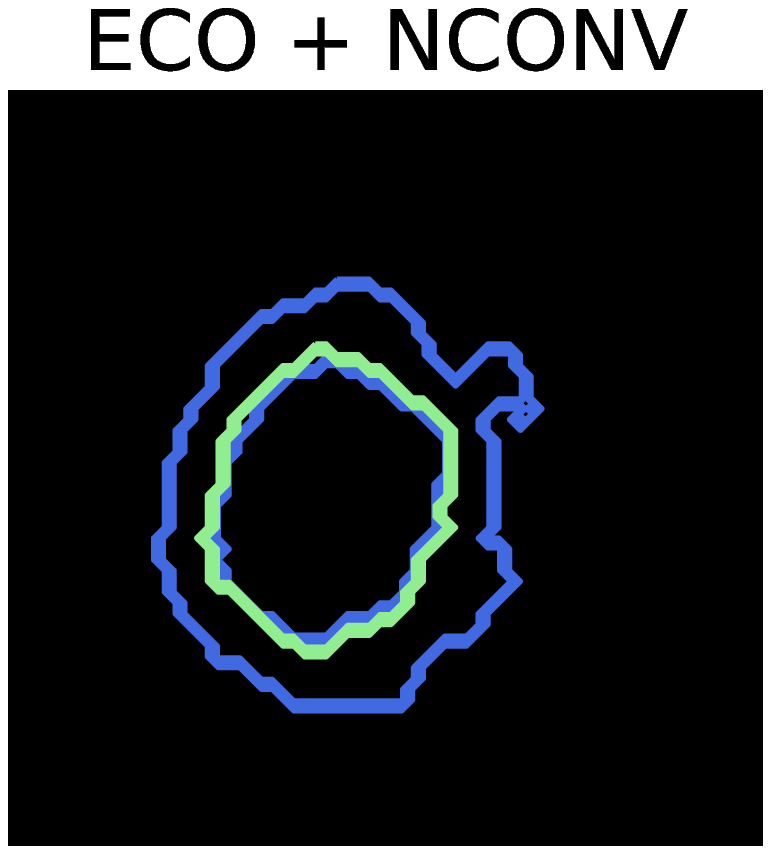}  
  \caption{}
  \label{fig:sub-eco_nconv}
\end{subfigure}
\begin{subfigure}{.19\textwidth}
  \centering
  \includegraphics[width=1.0\linewidth]{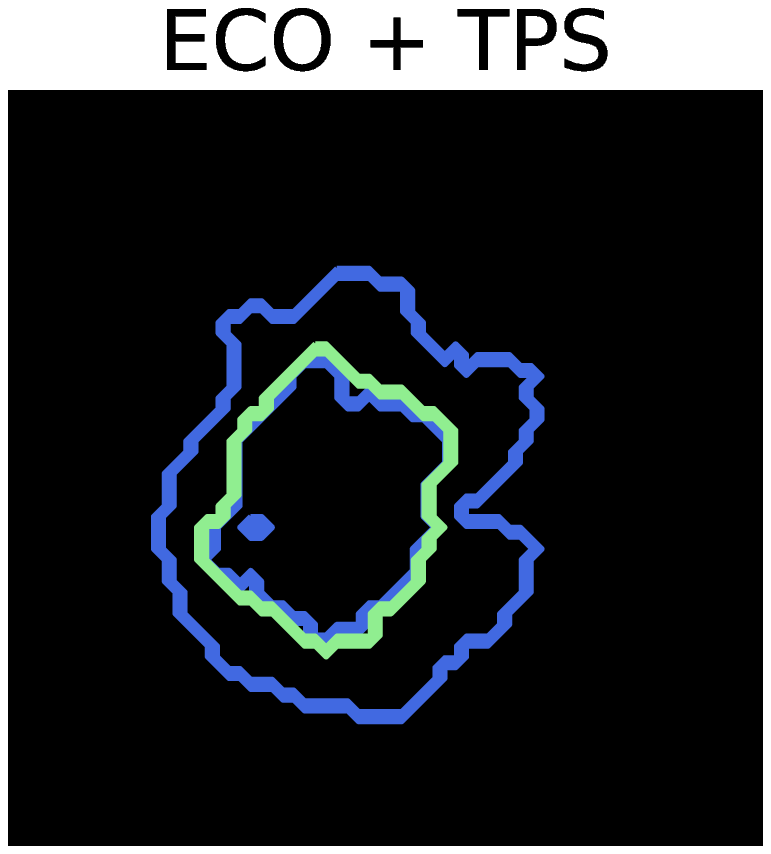}  
  \caption{}
  \label{fig:sub-eco_tps}
\end{subfigure}
\begin{subfigure}{.19\textwidth}
  \centering
  \includegraphics[width=1.0\linewidth]{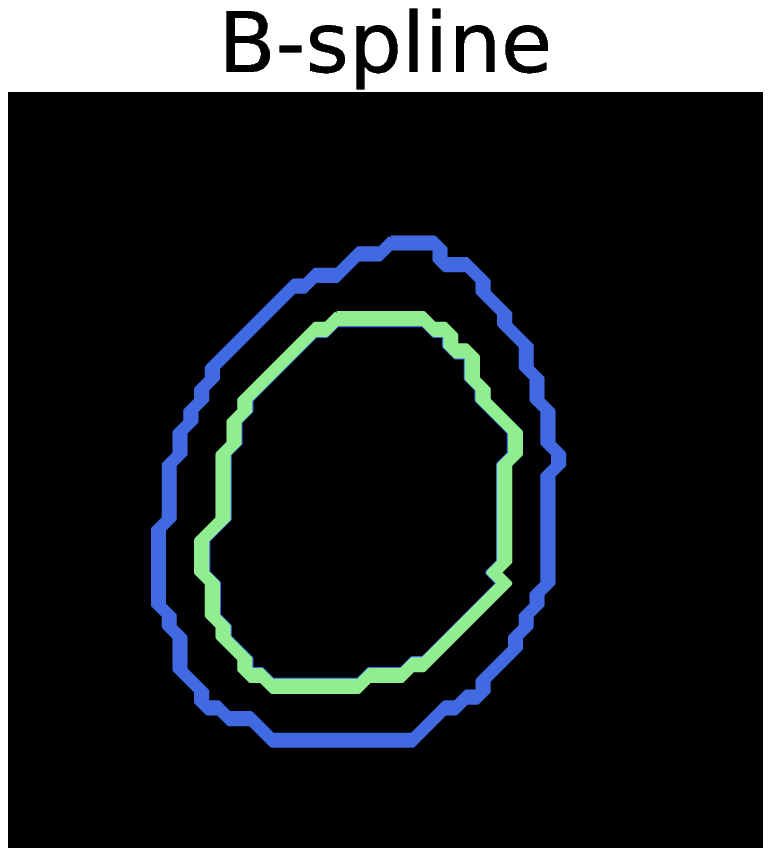}  
  \caption{}
  \label{fig:sub-bspline}
\end{subfigure}
\begin{subfigure}{.19\textwidth}
  \centering
  \includegraphics[width=1.0\linewidth]{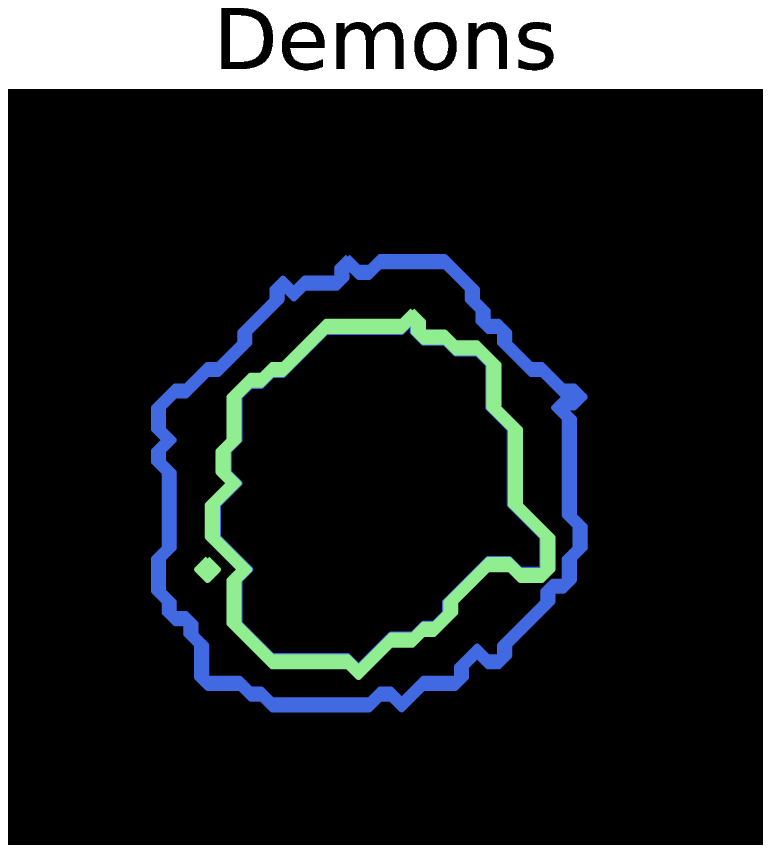}  
  \caption{}
  \label{fig:sub-demons}
\end{subfigure}
\caption{Illustration of the segmented areas. The ventricle (inside the green contour) and myocardium (between green and blue contour) segmentation areas are warped using the estimated displacement field for the different methods. Fig.~\subref{fig:sub-mov}~-~\subref{fig:sub-fixed} show the image at two time steps  ($t=0$ and $t=8$) for subject $15$. Fig.~\subref{fig:sub-manual}~-~\subref{fig:sub-demons} show the zoomed (yellow rectangle) segmented areas for the manual followed by the warped using the estimated displacement fields from time step $t=0$ to $t=8$.}
\label{fig:compare}
\end{figure}

In Fig.~\ref{fig:average} we show the mean Dice score at each time step, averaged across all subjects. In Fig.~\ref{fig:compare} we show the warped segmentation for each method between two time steps where the ventricle first is in its expanded position and then in its contracted position. This is illustrated by time steps $t=0$ and $t=8$ for subject $15$.
\section{Discussion and conclusion}
Our method shows an overall better performance in terms of Dice score when compared to conventional methods. As shown in Fig.~\ref{fig:average} and Fig.~\ref{fig:compare} our method handles larger movements better than conventional methods and it appears more clearly for smaller deformable region such as the myocardium. The method allows a trade-off between accuracy and computational capacity because of the flexibility in choosing the number and position of the tracking points. Although our current implementation is not fast enough for real-time estimations, we remain hopeful. A single tracker can be executed in real-time \cite{danelljan2017eco} but in the current implementation all trackers run sequentially. Since the trackers are independent of each other and thus the potential for significant speed improvements through parallellization exists. We also expect further gains from relating the tracking score to the confidence, and from tweaking the data augmentation for medical applications. 
\section*{Acknowledgement}
This research was partially supported by the 
\ifthenelse
{\boolean{anonymous}}
{***}
{\emph{Wallenberg AI, Autonomous Systems and Software Program (WASP)} funded by Knut and Alice Wallenberg Foundation}. The authors would like to thank \ifthenelse
{\boolean{anonymous}}
{***}{Dr. David Tilly} for helpful comments.

%
% ---- Bibliography ----
%
% BibTeX users should specify bibliography style 'splncs04'.
% References will then be sorted and formatted in the correct style.
%
\bibliographystyle{splncs04}
\bibliography{ref}

\begin{thebibliography}{10}
\providecommand{\url}[1]{\texttt{#1}}
\providecommand{\urlprefix}{URL }
\providecommand{\doi}[1]{https://doi.org/#1}

\bibitem{balakrishnan2019voxelmorph}
Balakrishnan, G., Zhao, A., Sabuncu, M.R., Guttag, J., Dalca, A.V.: Voxelmorph:
  a learning framework for deformable medical image registration. IEEE
  Transactions on medical imaging  \textbf{38}(8),  1788--1800 (2019)

\bibitem{bhat2019learning}
Bhat, G., Danelljan, M., Gool, L.V., Timofte, R.: Learning discriminative model
  prediction for tracking. In: Proceedings of the IEEE International Conference
  on Computer Vision. pp. 6182--6191 (2019)

\bibitem{bolme2010visual}
Bolme, D.S., Beveridge, J.R., Draper, B.A., Lui, Y.M.: Visual object tracking
  using adaptive correlation filters. In: Proceedings of the IEEE computer
  society conference on computer vision and pattern recognition. pp. 2544--2550
  (2010)

\bibitem{bookstein1989principal}
Bookstein, F.L.: Principal warps: Thin-plate splines and the decomposition of
  deformations. IEEE Transactions on pattern analysis and machine intelligence
  \textbf{11}(6),  567--585 (1989)

\bibitem{dalca2019unsupervised}
Dalca, A.V., Balakrishnan, G., Guttag, J., Sabuncu, M.R.: Unsupervised learning
  of probabilistic diffeomorphic registration for images and surfaces. Medical
  image analysis  \textbf{57},  226--236 (2019)

\bibitem{danelljan2018learning}
Danelljan, M.: Learning Convolution Operators for Visual Tracking. Ph.D.
  thesis, Link{\"o}ping University (2018)

\bibitem{danelljan2017eco}
Danelljan, M., Bhat, G., Shahbaz~Khan, F., Felsberg, M.: {ECO}: Efficient
  convolution operators for tracking. In: Proceedings of the IEEE conference on
  computer vision and pattern recognition. pp. 6638--6646 (2017)

\bibitem{danelljan2016beyond}
Danelljan, M., Robinson, A., Khan, F.S., Felsberg, M.: Beyond correlation
  filters: Learning continuous convolution operators for visual tracking. In:
  European conference on computer vision. pp. 472--488 (2016)

\bibitem{dice1945measures}
Dice, L.R.: Measures of the amount of ecologic association between species.
  Ecology  \textbf{26}(3),  297--302 (1945)

\bibitem{eldesokey2019confidence}
Eldesokey, A., Felsberg, M., Khan, F.S.: Confidence propagation through cnns
  for guided sparse depth regression. IEEE Transactions on pattern analysis and
  machine intelligence  (2019)

\bibitem{henriques2012exploiting}
Henriques, J.F., Caseiro, R., Martins, P., Batista, J.: Exploiting the
  circulant structure of tracking-by-detection with kernels. In: European
  conference on computer vision. pp. 702--715 (2012)

\bibitem{hua2018normalized}
Hua, J., Gong, X.: A normalized convolutional neural network for guided sparse
  depth upsampling. In: IJCAI. pp. 2283--2290 (2018)

\bibitem{klein2007evaluation}
Klein, S., Staring, M., Pluim, J.P.: Evaluation of optimization methods for
  nonrigid medical image registration using mutual information and {B}-splines.
  IEEE Transactions on image processing  \textbf{16}(12),  2879--2890 (2007)

\bibitem{knutsson1993normalized}
Knutsson, H., Westin, C.F.: Normalized and differential convolution. In:
  Proceedings of IEEE Conference on Computer Vision and Pattern Recognition.
  pp. 515--523 (1993)

\bibitem{miller2006geodesic}
Miller, M.I., Trouv{\'e}, A., Younes, L.: Geodesic shooting for computational
  anatomy. Journal of mathematical imaging and vision  \textbf{24}(2),
  209--228 (2006)

\bibitem{paganelli2018mri}
Paganelli, C., Whelan, B., Peroni, M., Summers, P., Fast, M., van~de Lindt, T.,
  McClelland, J., Eiben, B., Keall, P., Lomax, T., et~al.: Mri-guidance for
  motion management in external beam radiotherapy: current status and future
  challenges. Physics in Medicine \& Biology  \textbf{63}(22),  22TR03 (2018)

\bibitem{revaud2015epicflow}
Revaud, J., Weinzaepfel, P., Harchaoui, Z., Schmid, C.: Epicflow:
  Edge-preserving interpolation of correspondences for optical flow. In:
  Proceedings of the IEEE conference on computer vision and pattern
  recognition. pp. 1164--1172 (2015)

\bibitem{ronneberger2015u}
Ronneberger, O., Fischer, P., Brox, T.: U-net: Convolutional networks for
  biomedical image segmentation. In: International Conference on Medical image
  computing and computer-assisted intervention. pp. 234--241 (2015)

\bibitem{rueckert1999nonrigid}
Rueckert, D., Sonoda, L.I., Hayes, C., Hill, D.L., Leach, M.O., Hawkes, D.J.:
  Nonrigid registration using free-form deformations: application to breast
  {MR} images. IEEE Transactions on medical imaging  \textbf{18}(8),  712--721
  (1999)

\bibitem{schmid2000evaluation}
Schmid, C., Mohr, R., Bauckhage, C.: Evaluation of interest point detectors.
  International Journal of computer vision  \textbf{37}(2),  151--172 (2000)

\bibitem{shi1994good}
Shi, J., Tomasi, C.: Good features to track. In: 1994 Proceedings of IEEE
  conference on computer vision and pattern recognition. pp. 593--600 (1994)

\bibitem{Sotiras2013}
Sotiras, A., Davatzikos, C., Paragios, N.: Deformable medical image
  registration: A survey. IEEE Transactions on medical imaging  \textbf{32}(7),
   1153--1190 (2013)

\bibitem{stephens1988combined}
Stephens, M., Harris, C.: A combined corner and edge detector. In: Alvey vision
  conference. vol.~15 (1988)

\bibitem{thirion1998image}
Thirion, J.P.: {Image matching as a diffusion process: an analogy with
  Maxwell's demons}. {Medical Image Analysis}  \textbf{2}(3),  243--260 (1998)

\bibitem{xing2006overview}
Xing, L., Thorndyke, B., Schreibmann, E., Yang, Y., Li, T.F., Kim, G.Y.,
  Luxton, G., Koong, A.: Overview of image-guided radiation therapy. Medical
  Dosimetry  \textbf{31}(2),  91--112 (2006)

\end{thebibliography}

\end{document}